\documentclass[aps,prd,superscriptaddress,amsfonts,amssymb,amsmath,
               eqsecnum,nofootinbib,twocolumn,floatfix]{revtex4}

\usepackage[utf8]{inputenc}
\usepackage[T1]{fontenc}
\usepackage{mathptmx}
\usepackage{graphicx}
\usepackage{dcolumn}
\usepackage{bm}
\usepackage{mathtools}
\usepackage{amsmath}
\usepackage{amssymb}
\usepackage{ulem}     
\usepackage{siunitx}
\usepackage{chemformula}
\usepackage{indentfirst}

\begin{document}

\date{\today}
	
\title{Intraresonance frequency combs in Kerr microresonators}

\author{Andrei N. Danilin}
\email{danilin920@gmail.com} % corresponding author
\affiliation{Faculty of Physics, Lomonosov Moscow State University, Moscow 119991, Russia}
\affiliation{Russian Quantum Center, Moscow 121205, Russia}

\author{Timur R. Yunusov}
\affiliation{Russian Quantum Center, Moscow 121205, Russia}
\affiliation{Moscow Institute of Physics and Technology, Dolgoprudny 141701, Moscow Region, Russia}

\author{Ekaterina S. Vahnitskaya}
\thanks{These authors contributed equally to this work.}
\affiliation{Russian Quantum Center, Moscow 121205, Russia}
\affiliation{Moscow Institute of Physics and Technology, Dolgoprudny 141701, Moscow Region, Russia}

\author{Alexey P. Dushanin}
\thanks{These authors contributed equally to this work.}
\affiliation{Russian Quantum Center, Moscow 121205, Russia}
\affiliation{Moscow Institute of Physics and Technology, Dolgoprudny 141701, Moscow Region, Russia}

\author{Sanli Huang}
\affiliation{International Quantum Academy, Shenzhen 518048, China}
\affiliation{Hefei National Laboratory, University of Science and Technology of China, Hefei 230088, China}

\author{Zhenyuan Shang}
\affiliation{Hefei National Laboratory, University of Science and Technology of China, Hefei 230088, China}

\author{Junqiu Liu}
\affiliation{International Quantum Academy, Shenzhen 518048, China}
\affiliation{Hefei National Laboratory, University of Science and Technology of China, Hefei 230088, China}

\author{Anatoly V. Masalov}
\affiliation{Russian Quantum Center, Moscow 121205, Russia}
\affiliation{Lebedev Physical Institute, Russian Academy of Sciences, Moscow 119991, Russia}

\author{Dmitry A. Chermoshentsev}
\affiliation{Russian Quantum Center, Moscow 121205, Russia}
\affiliation{Moscow Institute of Physics and Technology, Dolgoprudny 141701, Moscow Region, Russia}

\author{Igor A. Bilenko}
\affiliation{Faculty of Physics, Lomonosov Moscow State University, Moscow 119991, Russia}
\affiliation{Russian Quantum Center, Moscow 121205, Russia}

\begin{abstract}
For more than 20 years, optical microresonators have served as the backbone of integrated nonlinear photonics, exploiting Kerr nonlinearity to generate octave-spanning frequency combs, enable quantum effects, and drive optical parametric oscillators. Since the inception of microresonator-based nonlinear optics, related studies have focused primarily on regimes in which photons with distinct resonant modes can interact. Although multiple comb lines can occupy a single resonance during the Kerr comb formation process, their mutual interactions have remained largely unexplored. Here we demonstrate a Kerr comb formation that is confined to a single resonance of a microresonator via dual-pumping. $\text{MHz}$-scale comb-line spacing reveals previously unobserved Kerr-comb dynamics, featuring parametrically driven phase multistability that can be observed directly in the temporal domain. Two laser pumps serve as phase-coupled references for heterodyne read-out, simplifying the measurements.
\end{abstract}

\maketitle

\section{Introduction}
Kerr-nonlinearity-induced frequency combs in high-$\mathrm{Q}$ microresonators have profoundly influenced fundamental metrology \cite{drake2019terahertz}, microwave photonics \cite{liang2015high, jost2015counting}, optical parametric oscillators (OPOs) \cite{savchenkov2004low, brodnik2025nanopatterned}, and photonic computing applications \cite{okawachi2016quantum, okawachi2020demonstration, yang2021squeezed}. At the heart of the comb generation process lies the parametric four-wave mixing (FWM) of lines at microresonator resonances, separated by their free spectral range (FSR). The dual-pumping of distinct microresonator resonances allows the generation of frequency combs \cite{PhysRevA.79.041805, Lucas2018, Wang2016, PhysRevA.90.013811} and degenerate optical parametric oscillation (DOPO) via FWM, in which a new signal appears precisely at the midpoint between two pump frequencies that are tuned to resonances separated by an even number of FSRs  \cite{okawachi2015dual, okawachi2016quantum, tatarinova2025optimizationdegenerateopticalparametric, PhysRevA.105.033524, PhysRevApplied.20.054036}. Such parametrically generated signals demonstrate several important properties: two stable phase states and quantum quadrature squeezing \cite{ulanov2025quadrature, zhang2021squeezed, zhao2020near}. These qualities are essential in optical quantum annealing algorithms \cite{PhysRevE.58.5355, okawachi2020demonstration, doi:10.1126/sciadv.abh0952} and universal quantum computations \cite{doi:10.1126/science.abe8770, AghaeeRad2025}.  

In this work we demonstrate a four-wave mixing process within the single resonance bandwidth of a microresonator. This effect can also be induced under single resonance dual-pumping, which results in generation of intermediate, equidistant spectral lines within the resonance bandwidth. Unlike Kerr combs with lines located in distinct cavity resonances \cite{Kerr_comb_del2007optical, herr2016dissipative}, in \textit{intraresonance Kerr combs} nonlinear interaction is confined to one resonance. The observed intraresonance frequency comb with $n$ equally spaced intermediate spectral lines is shown to be $(n+1)$-fold phase-stable. We observe states with an $n$ of up to $8$, opening a route from biphasic to multi-phase encoding. As a degenerate intraresonance signal ($n=1$) is parametrically induced, it is phase-bistable \cite{okawachi2016quantum}. These spectral lines manifest at the detector output as beats at the subharmonics of the dual-pump beat note. A two-mode intraresonance generation scheme ($n=2$) was previously predicted in atomic ensembles by Ref. \citep{Koch:89_Masalov}. 

We experimentally demonstrate control over these phase states using a weak external seed at one of the parametric frequencies. Similar phenomena have been reported under bichromatic pumping, where subharmonic spectral lines emerge and are been attributed to the breathing dynamics observed with cnoidal-wave modulation \cite{Gao:24}. We adapt existing numerical models to the proposed intraresonance dual-pumping approach and find good agreement with the experimental results. The pump-signal beat frequencies are in the tens-of-MHz range, enabling a direct time-domain analysis. The influence of dispersion can be neglected, eliminating the need for dispersion-engineering techniques. Moreover, at the normal dispersion level of the pumped microresonator the adjacent resonant modes remained unexcited.

\section{Results}
Kerr comb generation has been demonstrated on a variety of platforms, with $\mathrm{Si_3N_4}$ being the most mature of these platforms~\cite{Liu2021,Xuan:16,Ji:17}. We therefore fabricate and use integrated $\mathrm{Si_3N_4}$ devices to demonstrate the desired effect \cite{Ye:23}. We study phase multistability using a resonance in a Kerr-nonlinear $\mathrm{Si_3N_4}$ microresonator with normal group-velocity dispersion \cite{Luo2024, Sun2025}, a loaded linewidth of \SI{24}{\mega\hertz} and an FSR of \SI{199}{\giga\hertz}. This allows a low nonlinear threshold down to \SI{130}{\micro\watt}~\cite{PhysRevA.71.033804}. Although our dual-pump power may exceed this threshold by orders of magnitude, a Kerr comb does not form when the pump-resonance frequency offset is optimized for the intraresonance comb. However, both degenerate and nondegenerate interactions between pumps manifest in an intraresonance Kerr comb with a line spacing of $\Delta f/(n+1)$. The resonance linewidth imposes a limitation on the spectral envelope of the intraresonance comb analogously to the role of dispersion parameters in the conventional multimode single-pump Kerr comb regime~\cite{herr2014temporal}. In contrast with Kerr combs that are generated by dual-pumping distinct cavity resonances~\cite{Lucas2018,PhysRevA.79.041805,Wang2016}, where the number of comb lines between the pumps is limited by the number of intervening cavity modes, intraresonance dual-pumping imposes no such limit on $n$.

Because the wavevectors of the intraresonance comb are extremely close, dispersion \cite{herr2016dissipative, Moille2023} ceases to be the limiting factor with respect to energy redistribution; instead the intracavity dual-pump power limits the number of intervening spectral lines $n$. Therefore, the spectral shape of the resonance, pump power and pump-resonance frequency offsets are the main factors influencing the nonlinear spectrum. 

\begin{figure*}[tbp]
\includegraphics[width=\textwidth]{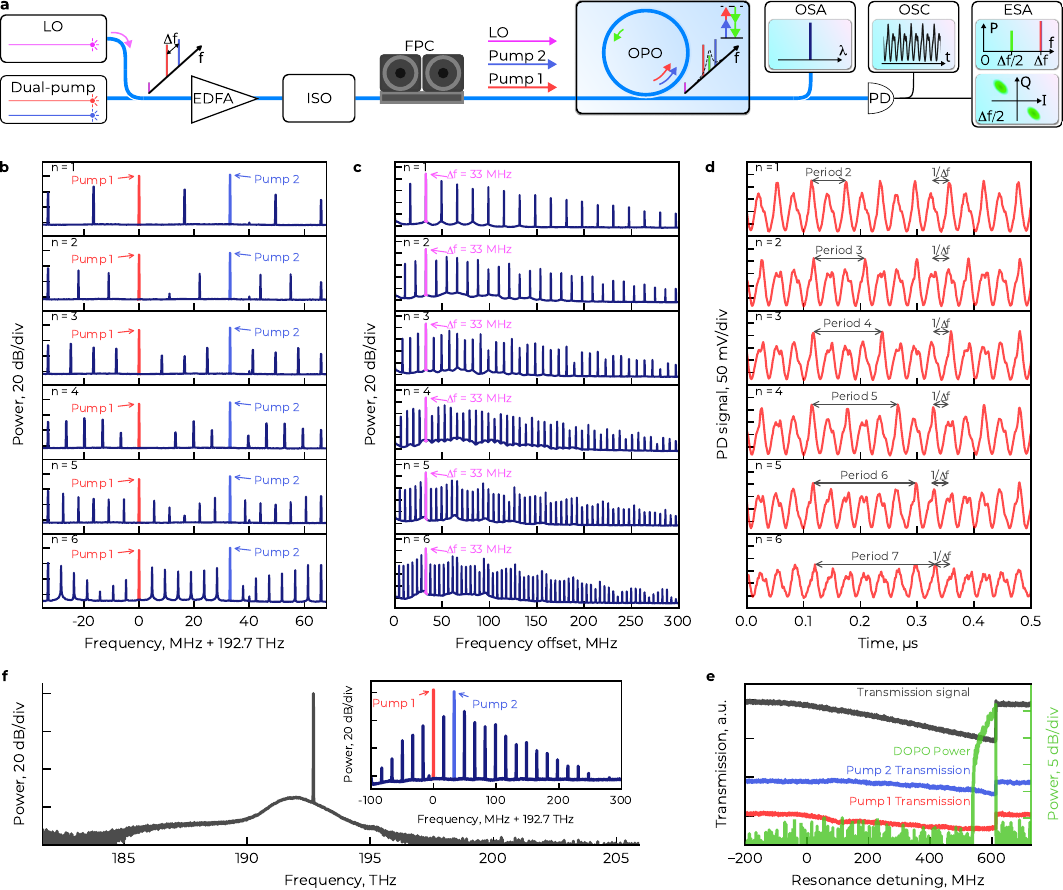}
\caption{ 
\textbf{a}: Experimental setup used to to generate multiple intraresonance subharmonics in a $\chi(3) $ microresonator. OSA, optical spectrum analyser; ESA, electrical spectrum analyser; OSC, oscilloscope; ISO, optical isolator; EDFA, erbium-doped fibre amplifier with an isolator; PD,  photodetector; FPC, fibre polarisation controller; LO, local oscillator. See the Methods section for a detailed description of the setup. 
\textbf{b}: Spectrum of the LO-comb beat on the PD, recorded on the ESA, for multiple values of $n$.
\textbf{c}:  Spectrum of the comb self-beat on the PD, recorded on the ESA, for multiple values of $n$.
\textbf{d}: Temporal intensity profile of the out-coupled field, recorded on the PD, for $n$ parametrically generated lines between pumps separated by $\Delta f=33~\mathrm{MHz}$.
\textbf{e}: Optical spectrum of the out-coupled field recorded on the OSA. Inset: Heterodyne probing with the LO reveals the fine intraresonance DOPO structure.
\textbf{f}: Oscillogram of the pump-resonance transmission versus the laser detuning from the cold resonance, showing the spike in DOPO power when both pumps are tuned to resonance.}
\label{fig:Fig_1_Scheme_Spectra}
\end{figure*}

A dual-pump source can be implemented using two individual lasers, a technique that has been proven to reliably yield Kerr frequency combs \cite{okawachi2015dual, okawachi2016quantum}. Although this method reveals intraresonance combs, it lacks the required relative phase stability for allowing precise phase measurements. Hence, we choose to modulate a single continuous-wave (CW) laser to generate a dual-pump source with extremely low relative phase noise \cite{ulanov2025quadrature}. Using the setup shown in Fig.~\ref{fig:Fig_1_Scheme_Spectra}a, we dual-pump the same $\chi^{(3)}$ resonance. This provides wide tunability relative to the resonance linewidth of $24~\mathrm{MHz}$. We set the pump separation to $\Delta f=33~\mathrm{MHz}$. Weak local oscillator (LO) serve as comb-envelope read-out. The power of the LO is set orders of magnitude below the power of the pump and the degree of its detuning from the resonance is sufficiently large to avoid nonlinear participation. All three optical fields are then amplified by an EDFA and coupled to the microresonator.

We tune the dual-pump across the resonance by adjusting the CW laser frequency. When both pump wavelengths are aligned with the resonance and the circulating power is sufficient, nonlinear generation occurs: new lines emerge within the resonance and are mutually coherent with the pumps (Fig.~\ref{fig:Fig_1_Scheme_Spectra}b). The temporal waveform is directly observed on an oscilloscope via a photodetector. The oscilloscope trace of the optical transmission, the individual pump powers and the DOPO power --- which highly resemble the parametrically driven pure-Kerr temporal solitons reported previously \cite{moille2024parametrically} --- are shown in Fig.~\ref{fig:Fig_1_Scheme_Spectra}c. The pump powers are inferred from their beat notes with the LO, while the total transmission is measured with the PD. When both pumps are in resonance, the power of the DOPO sharply increases at a specific pump offset (Fig.~\ref{fig:Fig_1_Scheme_Spectra}e). The optical spectrum analyser (OSA) records the out-coupled optical spectrum (Fig.~\ref{fig:Fig_1_Scheme_Spectra}f) and checks for sidebands in other resonances. Because the OSA cannot resolve the fine comb envelope, we heterodyne the optical field with the LO; the downconverted RF comb with a spacing of $\Delta f/2$ is shown in the inset of
Fig.~\ref{fig:Fig_1_Scheme_Spectra}f. The RF power spectra shown in Fig.~\ref{fig:Fig_1_Scheme_Spectra}c are measured on an electrical spectrum analyser (ESA) from the beat notes of the comb with itself. The intraresonance comb lines are equidistant; hence, with the known $\Delta f$ and the observed $n$, the beat frequencies between the parametric lines and the pumps are integer multiples of $\Delta f/(n+1)$. 

The value of $n$, which sets the spectral spacing $\Delta f/(n+1)$, depends on microresonator properties such as the resonance width, depth and splitting. Beyond these intrinsic properties, $n$ is controlled by experimentally accessible parameters: the intracavity pump power, the pump frequency spacing $\Delta f$ and the detuning from resonance. A complete mapping over the intrinsic microresonator properties and pump parameters is still lacking; nevertheless, the state selection scheme is predetermined. By adjusting the dual-pump power and the detuning one can choose the number of emission lines $n$ that populate the band between pumps (see Methods). For each case, knowing $n$ and therefore line spacing $\Delta f/(n+1)$, we demodulate the beat note signal and observe phase multistability and asymmetric quadrature amplification.

\begin{figure*}[ht]
\centering
\includegraphics[width=1.0\textwidth]{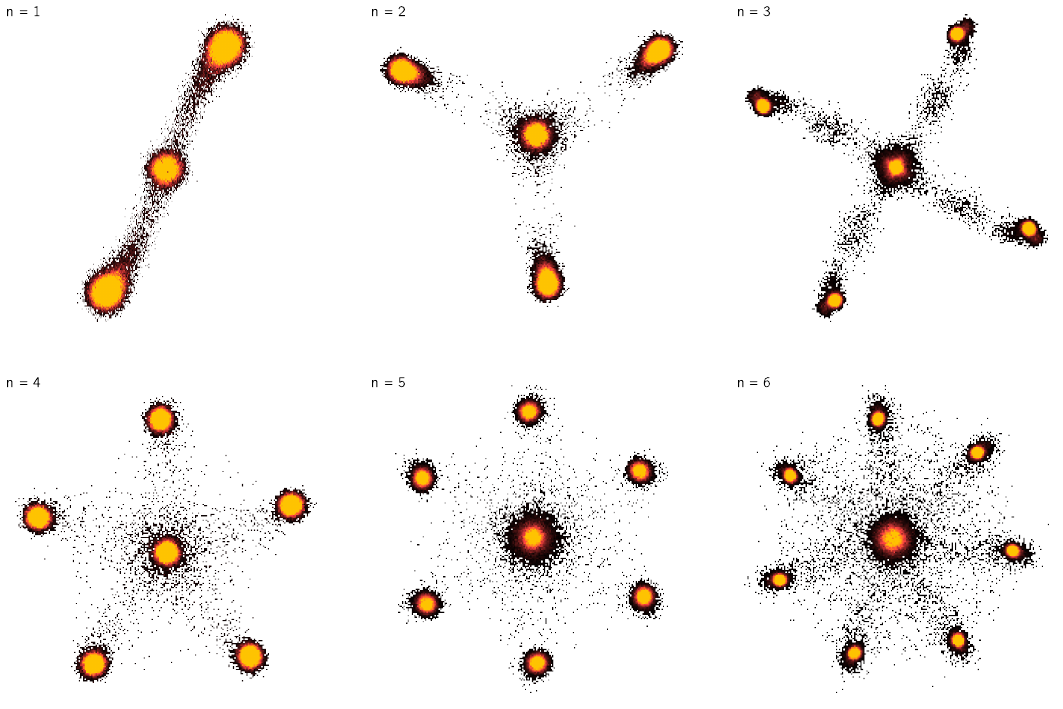}% Here is how to import EPS art
\caption{Quadrature-density (I/Q) diagrams of the PD electrical signal, recorded on an ESA for multiple values of $n$. The measurements are taken with pump modulation to toggle the system between above- and below-threshold operation and thereby reveal all possible phase realisations.}
\label{fig:IQ_histograms}
\end{figure*}

The corresponding time traces and RF spectra are shown in Figs.~\ref{fig:Fig_1_Scheme_Spectra}a--c. For each $n$, the time trace in Fig.~\ref{fig:Fig_1_Scheme_Spectra}c reflects a Turing pattern \cite{Turing_Patterns_PhysRevA.89.063814} formed within the cavity, yielding a stable and reproducible waveform. As $n$ increases, the repetition period also increases, reflecting the larger number of spectral lines within the cavity. In all cases, the waveform remains deterministic because of the strict phase relationships among the comb lines, as evidenced by the production of a clean beat note.

As indicated in a later section, the relative phases between adjacent comb lines are identical; hence, the phase can be retrieved from the comb self-beat. We demodulate the PD signal to recover the relative phases of all the parametrically generated lines and obtain complete information about the comb. When the beat note is subjected to I/Q demodulation, it generates multiple discrete stability points. This spontaneous phase-symmetry breaking is clearly visible in Fig.~\ref{fig:IQ_histograms}: each time the system is driven from below the threshold to above it, one of the allowable phase states is selected erratically, yielding the observed set of phase-stable points, which is consistent with Eq.~\eqref{eqn:phase_stability_points}.

\subsection{Controllable comb phase state switching} \label{sec_Switching}
\begin{figure*}[ht]
\includegraphics[width=1.0\linewidth]{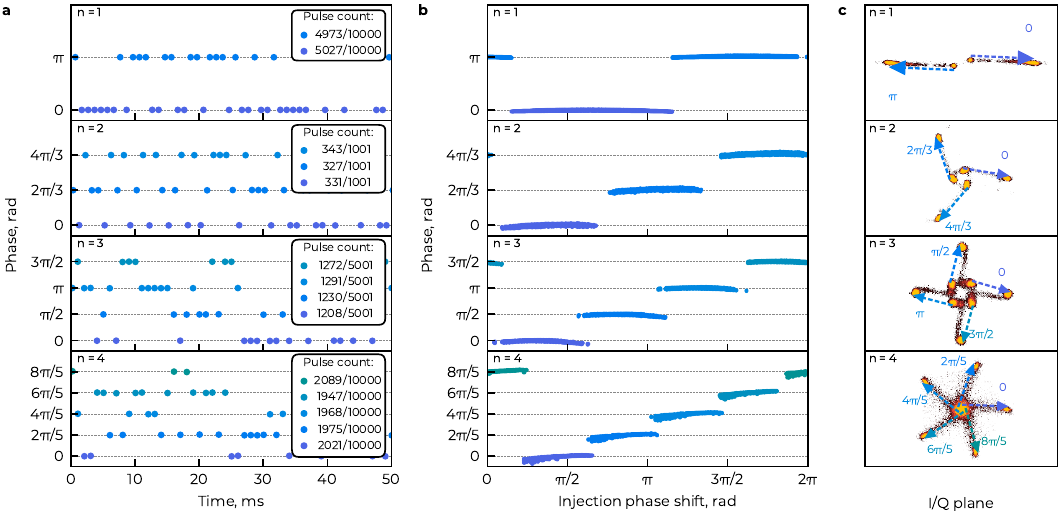}% Here is how to import EPS art
\caption{Injection measurements obtained with pump modulation toggling the generation process on and off.
\textbf{a}: Spontaneous symmetry breaking in the absence of external injection; each point marks the phase of an individual realisation.
\textbf{b}: Evolution of the generation phase while the injection phase is varied. The time axis is rescaled to the injection phase.
\textbf{c}: Quadrature phase diagram showing all possible phase states of the given regime as the injection phase is varied stepwise. Arrows indicate transitions from below- to above-threshold regimes.}
\label{fig:Fig_3_Injection}
\end{figure*}

We switch the generation process on and off by modulating the pump, forcing the system through the below-threshold region and initiating a new realisation each cycle. Without any injections (Fig.~\ref{fig:Fig_3_Injection}a), the oscillation phase for each realisation arbitrarily takes one of the allowable values given by Eq.~\eqref{eqn:phase_stability_points} for the observed $n$.

Although the spin comprises numerous optical lines, injection of a single, coherent, weak tone is sufficient. We seed the state at the frequency offset corresponding to one of the parametrically generated lines, $f_{p1}+\Delta f/(n+1)$. The seed follows the same optical path as the pumps do, ensuring relative phase stability. With coherent injection, the microresonator acts as a phase-sensitive parametric amplifier with $(n+1)$ favourable phase states. By shifting the injection phase through the phase shifting of the seed signal, we drive the oscillation between all allowed phase states (Figs.~\ref{fig:Fig_3_Injection}b,c). 

\subsection{Qualitative description model} \label{sec:Masalov_model}
Two pump waves in a microresonator fill the medium with a continuous spectral background because of the spontaneous parametric processes of type $f_{p1} + f_{p2} \Rightarrow f_{1}+ f_{2}$ and $2 f_p \Rightarrow f_{1}+ f_{2}$. If the intracavity power of the dual-pump setup is sufficient, the four-wave interactions favour numerous distinct frequency components of the continuous spectral background at frequencies between the interactions~\cite{okawachi2015dual, turitsyn2015inverse}, as shown experimentally in Fig.~\ref{fig:Fig_1_Scheme_Spectra}). A wave with a frequency at the centre between the pump waves has an obvious advantage because of the degeneracy of the FWM process $f_{p1} + f_{p2} \Rightarrow f_1$. In this case, the phase of the generated wave $\varphi_1$ is determined by the equations for the nonlinear transformation of the pump phases $\Phi_1$ and $\Phi_2$ \cite{klyshko2018photons}:
\begin{equation}
    2\varphi_1 = \Phi_{p1} + \Phi_{p2}+2\pi N,
    \label{eq:Phase_Matching_2st}
\end{equation}
with $N \in \mathbb {Z}$, yielding two phase-stable $\varphi_1$ solutions separated by $\pi$: $\varphi_1 = (\Phi_{p1} + \Phi_{p2})/2$ and $\varphi_1 = (\Phi_{p1} + \Phi_{p2})/2+\pi$. Bounded by the resonance linewidth and the intracavity optical power, this cascading process results in the formation of a Kerr frequency comb confined within a resonance.

These solutions manifest differently when the subharmonic beats $f_{p2}-f_{1} = f_{1}-f_{p1}$ of the beat frequencies of the two pumps $f_{p2}-f_{p1}$ are observed. If the beat phase at the fundamental frequency is determined by the phase difference  $\Phi_{p1} -  \Phi_{p2}$, the subharmonic phase has a value of $\Phi_{p2} -  \varphi_{1} = \varphi_{1} - \Phi_{p1} = (\Phi_{p1} + \Phi_{p2})/2 +\pi N$. Thus, when the central component is generated, subharmonic phase bistability occurs, as shown experimentally in Fig.~\ref{fig:IQ_histograms} and Fig.~\ref{fig:Fig_3_Injection}a. The phase relations discussed above do not take into account mismatches associated with the resonator's dispersion properties near the resonance line. The effect of mismatches can suppress the oscillation. It is likely that the dispersion-induced change in mismatches during pump frequency tuning leads to the dominance of oscillation at one of the frequency sets (see Fig.~\ref{fig:Fig_6_Map} in the Methods section).

With an increased parametric gain states with $n\in\mathbb{N}$ equidistant frequencies between the pumps can be generated. Each case is associated with $(n+1)$ solutions to the phase-matching condition equations. Which state actually arises depends on the intracavity pump power, the pump frequency offset and the resonance spectral shape. Generalizing to $n$ parametrically generated lines, the stable oscillation phase difference between adjacent comb lines $\Delta \varphi$ satisfies
\begin{equation}
    \Delta\varphi = \dfrac{2\pi}{n+1}N.
    \label{eqn:phase_stability_points}
\end{equation}

\subsection{Numerical modelling}
We numerically simulate the single-resonance dual-pumping of a Kerr microresonator with coupled-mode equations \cite{Herr2012, chembo2010modal, herr2014temporal} with a simulated pump power of \SI{5}{\milli\watt} and a frequency separation of $\Delta f = 27\,\text{MHz}$. We chose experimentally determined microresonator and pump parameters for this simulation.

\begin{figure*}[ht]
\includegraphics[width=0.9\linewidth]{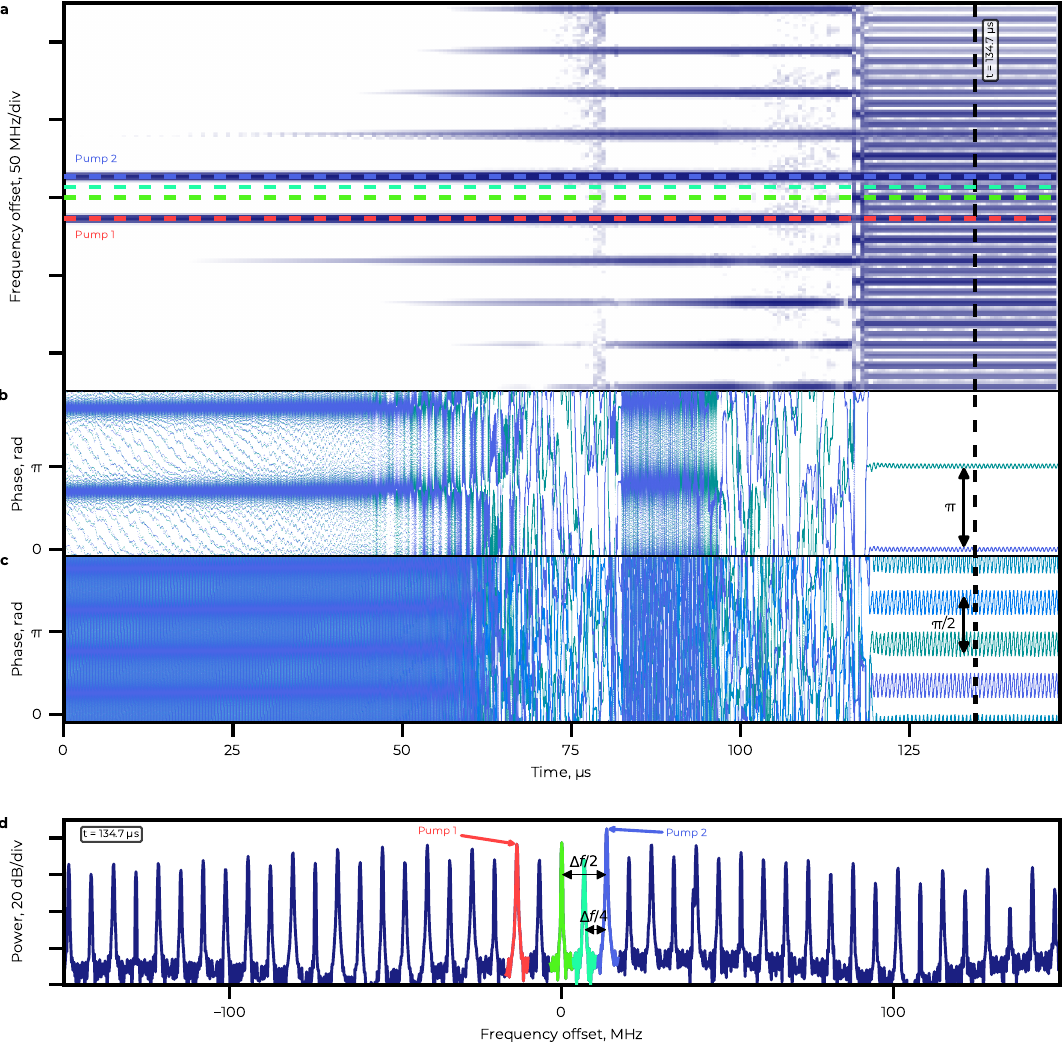}% Here is how to import EPS art
\caption{Simulation results regarding the intraresonance field of the microresonator under a bichromatic pumping process that is continuously tuned towards a cavity resonance. The pump-resonance detuning level is swept from \SI{-346}{\mega\hertz} to \SI{209}{\mega\hertz}, and the scan is halted at \SI{118.2}{\micro\second}. \textbf{a}: Spectrogram of the intracavity field during the detuning sweep. \textbf{b}: Phase bistability of the degenerate (DOPO) line at $\Delta f/2$ from each pump (green on \textbf{a,d}). The colours denote independent realizations. \textbf{c}: Fourfold phase stability of the line detuned by $\Delta f/4$ from the right pump (cyan in \textbf{a,d}). The colours denote independent realizations. \textbf{d} The
intracavity emission spectrum observed at \SI{134.7}{\micro\second}.}
\label{fig:2st_spec_plus_phase}
\end{figure*}

When the scan from the blue side to the red side of the resonance begins, the first \SI{60}{\micro\second} of the simulation exhibits a below-threshold state in which all $(n+1)$ phase states are simultaneously present, as seen in the computed spectrogram in Fig.~\ref{fig:2st_spec_plus_phase}a. As the pumps are tuned deeper into the same resonance, new intraresonance lines appear. The line at a $\Delta f/2$ offset from the pumps exhibits the expected phase bistability (Fig.\ref{fig:2st_spec_plus_phase}b), whereas the line at a $\Delta f/4$ offset exhibits the expected fourfold (quattro) stability (Fig.~\ref{fig:2st_spec_plus_phase}c). We stop the detuning sweep at \SI{118.2}{\micro\second} to allow the system to reach a steady state. The steady-state spectrum at \SI{134.7}{\micro\second} is shown in Fig.~\ref{fig:2st_spec_plus_phase}d, where a $n=3$ state is established. The central (degenerate) line, located at $\Delta f/2$ from either pump, exhibits phase bistability with a phase separation of $\pi/2$ between realizations (Fig.~\ref{fig:2st_spec_plus_phase}b). The line detuned by $\Delta f/4$ from the right pump exhibits fourfold phase stability, with a phase separation of $\pi/4$ between realizations (Fig.~\ref{fig:2st_spec_plus_phase}c). To reveal these multistable phase states, we repeat these otherwise identical detuning sweeps multiple times with independently initialized small perturbations; different realizations are indicated by different colours in the phase plots.

\section{Discussion}
We have identified and controlled a regime of intraresonance Kerr-comb formations in which dual-pumping drives the parametric oscillation process within a single resonance. These formations constitute distinct class of dissipative structures in microresonators: the comb spacing and phase organization schemes are governed not by intermodal FSRs and dispersion engineering but by the single-mode resonance profile together with the pump separation $\Delta f$ and the intracavity pump power levels.

Compared with conventional microcombs, the proposed mechanism involves relaxation of the usual dispersion constraints. Because all the participating fields lie within a single resonance, material and geometric dispersion play a reduced role in setting the spectrum. Instead, the resonance spectral shape acts as the primary spectral filter. This filtering approach suppresses parasitic channels and favours degenerate processes, making it similar to strategies that damp unwanted sidebands in photonic-molecule systems \cite{zhang2021squeezed, tomazio2024tunable}. The practical consequence is that the number of generated lines $n$ between the two pumps is not capped by the number of intervening modes. Instead, it is selectable through $\Delta f$, the detuning level and power, which are parameters that are straightforward to vary experimentally.

The intraresonance Kerr comb offers several advantages and exhibits a plethora of nonlinear effects. The pump-signal beat frequencies are in the tens-of-$\text{MHz}$ range, enabling direct time-domain analyses. At a normal dispersion level of the pumped microresonator the adjacent resonant modes remained unexcited.

Our observations also suggest opportunities in microwave photonics and sensing. The proximity of the phase-coupled pumps provides a natural phase reference for heterodyne measurements, whereas the DOPO process is known to generate quadrature squeezing in the subthreshold regime. Together, these findings motivate future experiments on sub-Poissonian heterodyne quantum sensing and on modulated Kerr microcombs in which intraresonance FWM is induced near each comb line to engineer multifrequency squeezed states, promising resources for Boson sampling protocols using photonic qudits.

The dynamics that we uncover bear a strong resemblance to breather-type Kerr combs \cite{Lucas2017, Matsko:12, Yu2017}. We hypothesize that intraresonance combs may be directly related to breather formation: broadband hyperparametric oscillations confined within a single resonance seed a coherent central line where the gain is maximal, which then cascades via four-wave mixing to generate the observed low-frequency comb. This idea could also account for the empirical proximity between the RF beat note and the laser-resonator detuning seen in breathing states \cite{Lucas2017}. Further study of intraresonance comb dynamics is therefore likely to elucidate the mechanisms behind the breather-type Kerr comb formation.

Although our experiments involve high-$Q$ $\mathrm{Si_3N_4}$ microring, the mechanism is general: any high-finesse resonator with Kerr nonlinearity and a sufficient intraresonance gain should exhibit the same physics. Taken together, these attributes indicate that intraresonance Kerr combs could serve as compact, controllable building blocks for integrated photonic computations and multifrequency quantum resources.

\begin{figure*}[ht]
\includegraphics[width=1.0\linewidth]{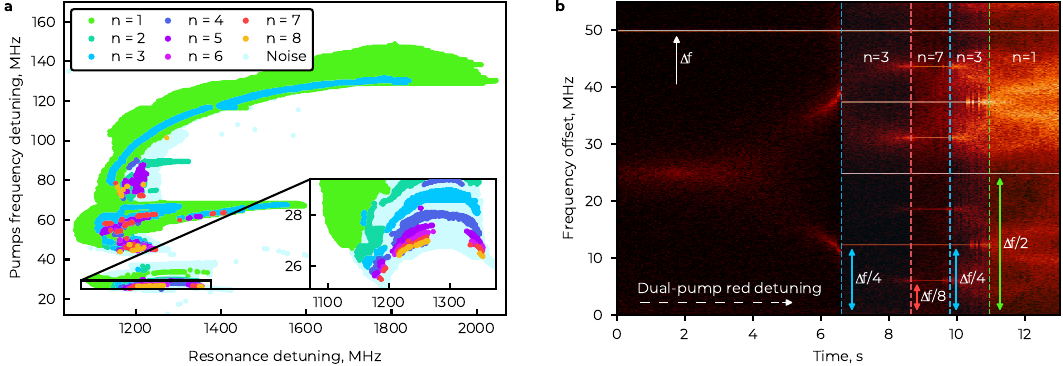}
\caption{
\textbf{a}: Map of the state realizations as a function of the dual-pump frequency detuning level $\Delta f$ and the degree of pump resonance detuning from the cold resonance under equal pump powers. Each data point corresponds to a specific oscillation state.
\textbf{b}: The time-frequency spectrogram of the photodetector signal as the dual-pump is red-detuned across the microresonator resonance. The observed sequence of intraresonance states is $n=3 \rightarrow 7 \rightarrow 3 \rightarrow 1$. The $n=7$ state arises from the four-wave mixing of the $n=3$ comb lines acting as pumps.}
\label{fig:Fig_6_Map}
\end{figure*}

\section{Methods}\label{sec_Methods}
The target $n$ values are empirically obtained by selecting the pump-resonance detunings and the pump powers. We map the accessible regimes (Fig.~\ref{fig:Fig_6_Map}a) by varying the pump separation $\Delta f$ and the detuning level relative to the resonance. The maps reveal broad regions that are associated with particular stability types. Regions with DOPO ($n=1$) are the most prominent type, which is consistent with their lower detection threshold of approximately \SI{0.4}{\milli\watt}. Depending on the pump power, a minimum pump separation $\Delta f$ is required to achieve generation between the pumps. For the studied resonance, this minimum is at least \SI{10}{\mega\hertz} and increases with increasing pump power. Increasing the pump power can also increase the values of $n$ (e.g., $n=7,8$, or higher), as shown in Fig.~\ref{fig:Fig_6_Map}a.

Some states sequentially arise from preexisting states: the comb lines of an initially formed state act as pumps and, via four-wave mixing, generate additional components. Such sequential states are typically observed within the broad $n=1$ and $n=2$ regions, where new oscillations are created from neighbouring lines of the preexisting state. For example, a sequential $n=5$ state can appear within either the $n=1$ or $n=2$ region: it arises when an $n=2$ state is generated within an existing $n=1$ state or, conversely, when an $n=1$ state is generated within an existing $n=2$ state. For example, in Fig.~\ref{fig:Fig_6_Map}b, after the $n=3$ state is established, continuous tuning can lead to the $n=7$ state, in which a DOPO is generated between the lines of the $n=3$ frequency comb. The additional lines exhibit phase bistability relative to the $n=3$ comb when the system remains on the same-generation branch. However, when the detuning level is swept from the blue side to the red side and both the $n=3$ and $n=7$ oscillations are re-established, the composite state exhibits eightfold phase stability. The selected regions are labelled "Noise", indicating a broadband parametric amplification background.

\section*{Author contribution}
A.N.D. suggested and implemented the intraresonance dual pumping. A.N.D., T.R.Y., E.S.V., and A.P.D. designed, built, and carried out the experiment. A.P.D., and E.S.V. analysed the data. A.V.M. developed the theoretical model of the effect. D.A.C. created the numerical model. A.N.D., T.R.Y. and D.A.C. carried out numerical simulations. S. H., Z. S. and J. L.  designed and fabricated the $\mathrm{Si_3N_4}$ chip devices. A.N.D. wrote the manuscript with input from all authors. D.A.C. and I.A.B. supervised the project.

\section*{Funding}
This work is supported by the Russian Science Foundation (project 23-42-00111). J.L. acknowledges support from the National Natural Science Foundation of China (Grant No.12261131503), Innovation Program for Quantum Science and Technology (2023ZD0301500), and Shenzhen Science and Technology Program (Grant No. RCJC20231211090042078). 

\section*{Acknowledgments}
A.N.D. acknowledge the Foundation for the Advancement of Theoretical Physics and Mathematics "BASIS" for personal support. 

% \bibliographystyle{apsrev} 
% \bibliography{sn-bibliography}

\end{document}